\documentstyle[12pt,epsf]{article}

\renewenvironment{thebibliography}[1]
{\normalsize
 \begin{list}{[\arabic{enumi}]}
 {\usecounter{enumi} \setlength{\parsep}{0pt}
  \setlength{\itemsep}{3pt} \settowidth{\labelwidth}{[#1]}
  \sloppy}}
{\end{list}}

\begin{document}

\hfill\vbox{\baselineskip14pt
            \hbox{\bf KEK-TH-443}
            \hbox{\bf KEK Preprint 95-80}
            \hbox{\bf\today}
            \hbox{H}}
\baselineskip20pt

\vspace*{1cm}
\begin{center}
{\Large{\bf Probing the Top-Quark Electric Dipole Moment\\
            at a Photon Linear Collider}}
\end{center}
\vspace*{1cm}

\begin{center}
\large S.Y. Choi and K. Hagiwara
\end{center}
\begin{center}
{Theory Group, KEK, Tsukuba, Ibaraki 305, Japan }
\end{center}
\vspace*{2cm}

\begin{center}
\large Abstract
\end{center}
\vspace{0.2cm}
\begin{center}
\begin{minipage}{14cm}
\baselineskip=20pt
\noindent
We probe the top-quark electric dipole moment (EDM)
in top-quark pair production via photon-photon fusion at
a photon linear collider. We show how linearly-polarized photon
beams can be used to extract information on the top-quark EDM
without the use of complicated angular
correlations of top-quark decay products.
If the luminosity of the laser back-scattered
photon-photon collisions is
comparable to that of the $e^+e^-$ collisions,
then the measurement of the top-quark EDM
obtained by counting top-quark-production events in
photon fusion can be as accurate as the measurement obtained by
studying the $t\bar{t}$ decay correlations
in $e^+e^-$ collisions using a perfect detector.
\end{minipage}
\end{center}
\vfill

\baselineskip=20pt
\normalsize

\newpage
\setcounter{page}{2}

A detailed study of top-quark physics is one of the most important
tasks of a linear $e^+e^-$ collider.
The fact that the top quark has a much larger mass than the other
quarks and leptons suggests that it may be susceptible to
effects from physics beyond the Standard Model (SM), to which
lighter quarks and leptons are not sensitive.

Detailed studies have been performed mainly in the processes
$e^+e^-\rightarrow t\bar{t}$\cite{Saa,Hawaii} including general
studies of $tt\gamma$, $ttZ$ and $tbW$ couplings\cite{KLY,AA,BN,ABB}.
At a photon linear collider we can measure the $tt\gamma$ and
$tbW$ couplings.
In this paper we investigate the possibility of extracting
the effective couplings of the top quark to the photon
by employing linearly polarized photons generated by the Compton
back-scattering of linearly polarized laser light on
electron/positron beams of linear $e^+e^-$ (or $e^-e^-$) colliders.

One may employ two methods to extract the top-quark effective
couplings at a photon linear collider.
One method makes use of the quasi-freely decaying property
of the top quark by measuring various spin correlations in the
$t\bar{t}$ final system, $(bW^+)(\bar{b}W^-)$ or
$(bf_1\bar{f}_2)(\bar{b}f_3\bar{f}_4)$.
The other method is to employ linearly-polarized photon beams
to measure various polarization asymmetries of the initial
states. It is, of course, possible to combine the two
methods. The former technique is essentially the same as
that employed in $e^+e^-$ collisions\cite{KLY,AA,BN,ABB}
with one difference; in $e^+e^-$ collisions the spin of the
$t\bar{t}$ system is restricted to $J=1$, while in photon
fusion $J=0$ or $J\geq 2$ is allowed.
Additionally it is easy to produce linearly polarized photon
beams through the Compton back-scattering of polarized laser
light off the initial electron/positron beams.
Hence a $\gamma\gamma$ collider provides some unique
opportunities.
We will show that CP-odd correlations can be measured by a
counting experiment where no information on the momenta and
polarization of the top quark decay products are needed.

A CP-odd asymmetry can be formed at a $\gamma\gamma$-collider
when the following $J=0$ amplitudes of two photons in the
CP-even and CP-odd states are both non-vanishing;
\begin{eqnarray}
{\cal M}\left[\gamma\gamma\rightarrow X({\rm CP}=+)\right]
 \propto \vec{\epsilon}_1\cdot\vec{\epsilon}_2,\qquad
{\cal M}\left[\gamma\gamma\rightarrow X({\rm CP}=-)\right]
 \propto (\vec{\epsilon}_1\times\vec{\epsilon}_2)\cdot
         \vec{k}_1,
\end{eqnarray}
where $\epsilon_1$ and $\epsilon_2$ are the polarizations of
the two colliding photons  and $\vec{k}_1$ is the momentum
vector of one photon in the $\gamma\gamma$ c.m. frame.
This fact has been used in Ref.~\cite{GG} to show that a Higgs
boson ($X=H$) of mixed CP character could be distinguished from a
pure CP eigenstate by the fact that a certain asymmetry between
production cross sections initiated by colliding photons of
different helicities is non-zero. A  similar analysis can
be applied to the $t\bar{t}$ production process ($X=t\bar{t}$)
where, for instance, the top-quark EDM generates CP-odd terms.
It is evident that the determination of the CP of $\gamma\gamma$
system requires linearly polarized colliding photons.

In general, a linearly polarized photon state is a superposition
of two helicity states with equal weight. The state vector of
a linearly polarized photon with the polarization along
an azimuthal angle $\phi$ in a given coordinate system can be
written as
\begin{eqnarray}
|\phi\rangle =\frac{1}{\sqrt{2}}
     \left[-e^{-i\phi}|\lambda=+\rangle
           +e^{i\phi}|\lambda=-\rangle\right].
\end{eqnarray}
The state vector of the two-photon system, where two photons
collide head-on, is the direct product
\begin{eqnarray}
|\phi_1,\phi_2\rangle
     =
\frac{1}{2}\left[e^{-i(\phi_1-\phi_2)}|++\rangle
                -e^{-i(\phi_1+\phi_2)}|+-\rangle
                -e^{i(\phi_1+\phi_2)}|-+\rangle
                +e^{i(\phi_1-\phi_2)}|--\rangle\right].
\end{eqnarray}
Then the scattering amplitude from the linearly polarized state to a
final state $X$ is
\begin{eqnarray}
\langle X |M|\phi_1,\phi_2\rangle
  =\frac{1}{2}
   \left[e^{-i(\phi_1-\phi_2)}M_{++}-e^{-i(\phi_1+\phi_2)}M_{+-}
        -e^{i(\phi_1+\phi_2)}M_{-+}+e^{i(\phi_1-\phi_2)}M_{--}\right].
\end{eqnarray}
Here we have used for helicity amplitudes the notation
\begin{eqnarray}
M_{\lambda_1\lambda_2}=\langle X|M|\lambda_1\lambda_2\rangle.
\end{eqnarray}
The azimuthal angle difference, $\chi=\phi_1-\phi_2$, is not
changed with respect to rotation along the two-photon direction,
while the azimuthal angle sum, $\varphi=\phi_1+\phi_2$, can be
cancelled by an overall rotation along the $z$-axis.
After taking an average over the azimuthal angle $\varphi$ we
obtain the matrix element squared as
\begin{eqnarray}
\Sigma(\chi)
 =\frac{1}{4}\sum_{\lambda_1,\lambda_2=\pm}|M_{\lambda_1\lambda_2}|^2
 -\frac{1}{2}\cos(2\chi){\rm Re}\left[M_{++}M^*_{--}\right]
 -\frac{1}{2}\sin(2\chi){\rm Im}\left[M_{++}M^*_{--}\right],
\end{eqnarray}
which differs from the unpolarized cross section by the last
two terms. Only the last term is CP-odd.

We now turn to the specific process $\gamma\gamma\rightarrow
t\bar{t}$ with Compton back-scattered laser lights as a photon
source at a future linear $e^+e^-$ ($e^-e^-$) collider.
The effective Lagrangian for the top-quark--photon
interaction is generalized by including terms of dimension five:
\begin{eqnarray}
{\cal L}_M=eQ_t\left(\frac{d_t}{2m_t}\right)
           \bar{t}\sigma^{\mu\nu}F_{\mu\nu}t,\qquad
{\cal L}_E=-ieQ_t\left(\frac{\tilde{d}_t}{2m_t}\right)
           \bar{t}\sigma^{\mu\nu}\gamma_5F_{\mu\nu}t,
\end{eqnarray}
where $\sigma^{\mu\nu}=\frac{i}{2}\left[\gamma^\mu,\gamma^\nu\right]$,
$F_{\mu\nu}$ is the photon field strength and $Q_t=2/3$ is
the top-quark electric charge in units of the proton charge $e$.
The coefficient $\mu_t=ed_tQ_t/2m_t$
is then called the anomalous magnetic dipole moment (MDM) of the
top quark, which receives one-loop contributions of the order
$\alpha_s/\pi m_t$.
This would very likely mask any new-physics contribution to the MDM.
On the other hand, the top-quark EDM,
$\tilde{\mu}_t=e\tilde{d}Q_t/2m_t$, violates T-invariance.
Its presence can hence be tested unambiguously via CP-odd
correlations.  Therefore, we focus in this letter on the measurement
of $\tilde{d}_t$.

The top-quark EDM modifies
the SM $\gamma t\bar{t}$ vertex as
\begin{eqnarray}
\Gamma^{\mu}(k,p,p^\prime)
     &=&-ieQ_t\left[\gamma^\mu+\tilde{d}_t\sigma^{\mu\nu}
        \gamma_5\frac{k_\nu}{2m_t}\right],
\end{eqnarray}
where $k=p-p^\prime$ is the four-momentum of the photon and
$p$ and $p^\prime$ are four-momenta of the incoming and outgoing
top quarks, respectively. With this modification,
the helicity amplitudes for the process
$\gamma\gamma\rightarrow t\bar{t}$ are
in the $\gamma\gamma$ c.m. frame given by
\begin{eqnarray}
{\cal M}_{\lambda_1\lambda_2;\sigma\bar{\sigma}}
        =
\frac{4\pi\alpha Q^2_tN_c}{(1-\hat{\beta}^2\cos^2\theta)}
A_{\lambda_1\lambda_2;\sigma\bar{\sigma}},
\end{eqnarray}
where
\begin{eqnarray}
&&A_{\lambda\lambda;\sigma\sigma}=-\frac{\sqrt{\hat{s}}}{2m_t}
  (\lambda+\sigma\hat{\beta})\left[\frac{4m_t^2}{\hat{s}}+
  \tilde{d}_t\left(i\lambda-\frac{\tilde{d}_t}{2}\right)
  (1+\lambda\sigma\hat{\beta}\cos^2\theta)\right],\nonumber\\
&&A_{\lambda\lambda;\sigma,-\sigma}=
  \tilde{d}_t\left(i\lambda-\frac{\tilde{d}_t}{2}\right)
  \hat{\beta}\cos\theta\sin\theta,\nonumber\\
&&A_{\lambda,-\lambda;\sigma\sigma}=\frac{\sqrt{\hat{s}}}{2m_t}
  \sigma\hat{\beta}\sin^2\theta
  \left[\frac{4m_t^2}{\hat{s}}-i\tilde{d}_t\sigma\hat{\beta}
  +\frac{\tilde{d}_t^2}{2}\right],\nonumber\\
&&A_{\lambda,-\lambda;\sigma,-\sigma}=\hat{\beta}
  \left(\lambda\sigma+\cos\theta\right)\sin\theta
  \left[1-\frac{\hat{s}}{8m_t^2}\tilde{d}_t^2
  (1-\lambda\sigma\hat{\beta}^2\cos\theta)\right],
\end{eqnarray}
where $\lambda,\bar{\lambda}$ and $\sigma/2,\bar{\sigma}/2$
are the two-photon and $t,\bar{t}$ helicities, respectively.
$\hat{s}$ is the $\gamma\gamma$ c.m.
energy squared, $\hat{\beta}=\sqrt{1-4m^2_t/\hat{s}}$, and
$\theta$ is the polar angle between the photon beam and
the top-quark directions.
The differential cross section of the process
$\gamma\gamma\rightarrow t\bar{t}$ is then given by
\begin{eqnarray}
\frac{{\rm d}\sigma}{{\rm d}\cos\theta}
     (\lambda_1\lambda_2;\lambda^\prime_1\lambda^\prime_2)
  & = &
 \frac{4\pi\alpha^2}{\hat{s}}
 \frac{Q^4_tN_c\beta}{(1-\hat{\beta}^2\cos^2\theta)^2}
 T_{\lambda_1\lambda_2;\lambda^\prime_1\lambda^\prime_2},\\
T_{\lambda_1\lambda_2;\lambda^\prime_1\lambda^\prime_2}
 & = &\sum_{\sigma_1\sigma_2=\pm}
  A_{\lambda_1\lambda_2;\sigma_1\sigma_2}
  A^*_{\lambda^\prime_1\lambda^\prime_2;\sigma_1\sigma_2}.
\end{eqnarray}
We find to first order in $\tilde{d}_t$,
\begin{eqnarray}
\frac{1}{4}\sum_{\lambda_1\lambda_2=\pm}
             T_{\lambda_1\lambda_2;\lambda_1\lambda_2}
 \!\! &=&\!\! \frac{1}{2}
    \left[1+2\hat{\beta}^2
   -(1+\sin^4\theta)\hat{\beta}^4\right],\nonumber\\
{\rm Re}[T_{++;--}]=-2(1-\hat{\beta}^2)^2,& &\qquad
  {\rm Im}[T_{++;--}]=-4(1-\hat{\beta}^2){\rm Re}(\tilde{d}_t).
\label{Matrix}
\end{eqnarray}
One can see from these expressions that, to first order,
the electric-dipole-form-factor, $\tilde{d}_t$, does not
contribute to the unpolarized cross-section, nor to the CP-even
term, ${\rm Re}[T_{++;--}]$, but it can contribute
to the CP-odd term, ${\rm Im}[T_{++;--}]$.
It is now clear that its measurement requires linear polarization
of the two photon beams, which can be achieved efficiently via
the Compton backscattering of laser beams\cite{GKS}.

The event rate for $t\bar{t}$ production can now be obtained by
folding a photon luminosity spectral function,
${\rm d}L_{\gamma\gamma}$,
with the $t\bar{t}$ production cross section as
\begin{eqnarray}
{\rm d}N={\rm d}L_{\gamma\gamma}\sum_{i,j=0}^{3}
         \langle \zeta_i\bar{\zeta}_j\rangle_\tau
         \sigma_{ij},
\end{eqnarray}
where $\zeta_i(\bar{\zeta}_i)$
are the Stokes parameters (with $\zeta_0=\bar{\zeta}_0=1$) for
$\gamma_1(\gamma_2)$, and $\sigma_{ij}$ are the corresponding
$t\bar{t}$ production cross-sections.
The linear-polarization transfer from the laser photons to
the high-energy photons is described by the $\zeta_1$ and
$\zeta_3$ components of the Stokes
vector, $\vec{\zeta}=(\zeta_1,\zeta_2,\zeta_3)$.
The length, $l=\sqrt{\zeta^2_1+\zeta^2_3}$, of these two
components depends on the final state
photon energy and on the value of the parameter
$x=2p_e\cdot p_\gamma/m^2_e$.
${\rm d}L_{\gamma\gamma}$ and
$\langle \zeta_i\bar{\zeta}_j\rangle_\tau$
are obtained as functions of the fraction $\tau(=\hat{s}/s)$
of the $\gamma\gamma$ c.m. energy-squared, $\hat{s}$, to the
$e^+e^-$ c.m. energy squared $s$ by averaging over collision,
including a convolution over the energy spectra for the
colliding photons.
Operation below the threshold for $e^\pm$ pair production
in collisions between the laser beam and the Compton-backscattered
photon beam requires $x\leq 2(1+\sqrt{2})$;
the lower bound on $x$ depends on the lowest available laser
frequency and the production threshold of the final state.
The linear-polarization transfer is large for small
values of $x$ if the photon energy to electron beam energy,
$y$, is close to the maximum value, i.e.
\begin{eqnarray}
y\rightarrow y_m=\frac{x}{1+x},
\label{ymax}
\end{eqnarray}
for a given parameter $x$. The photon energy spectrum is
given by
\begin{eqnarray}
\phi_0(y)=\frac{1}{1-y}+1-y-4r(1-r),
\end{eqnarray}
and the Stokes linear component, $l(y)=\phi_3(y)/\phi_0(y)$,
follows from
\begin{eqnarray}
\phi_3(y)=2r^2,
\end{eqnarray}
with $r=y/x(1-y)$.
The maximum value of $l(y)$ is reached at $y=y_m$,
\begin{eqnarray}
l(y_m)=\frac{2(1+x)}{1+(1+x)^2},
\end{eqnarray}
and approaches unity for small values of $x$.
Since only part of the laser polarization is transferred
to the high-energy photon beam, it is useful,
for the sake of discussion, to define the polarization
asymmetry ${\cal A}$ as
\begin{eqnarray}
{\cal A}(\tau)=\frac{\langle \phi_3\phi_3\rangle_\tau}{\langle
               \phi_0\phi_o\rangle_\tau},
\end{eqnarray}
where
\begin{eqnarray}
\langle \phi_i\phi_i\rangle_\tau
 =
 \frac{1}{{\cal N}^2(x)}\int^{y_m}_{\tau/y_m}\frac{{\rm d}y}{y}
 \phi_i(y)\phi_i(\tau/y).
\end{eqnarray}
The normalization factor ${\cal N}(x)$ is defined as
\begin{eqnarray}
{\cal N}(x)=\int^{y_m}_0\phi_0(y){\rm d}y
    ={\rm ln}(1+x)\left[1-\frac{4}{x}-\frac{8}{x^2}\right]
      +\frac{1}{2}+\frac{8}{x}-\frac{1}{2(1+x)^2}.
\end{eqnarray}

For the linearly-polarized laser beams the number of the $t\bar{t}$
events is given by
\begin{eqnarray}
{\rm d}N=\kappa^2L_{ee}\langle \phi_0\phi_0\rangle_\tau{\rm d}\tau
     \left[\sigma_0-\cos(2\chi){\cal A}\sigma_{A_1}
          -\sin(2\chi){\cal A}\sigma_{A_2}{\rm Re}(\tilde{d}_t)\right],
\label{Number}
\end{eqnarray}
where for the two-photon luminosity function the relation
\begin{eqnarray}
{\rm d}L_{\gamma\gamma}=\kappa^2L_{ee}
                       \langle \phi_0\phi_0\rangle_\tau{\rm d}\tau
\label{Lum_rel}
\end{eqnarray}
is used and the respective cross sections are given by
\begin{eqnarray}
   \sigma_0(\hat{s})
&=&\frac{4\pi\alpha^2Q^4_tN_c\hat{\beta}}{\hat{s}}
   \left[-2+\hat{\beta}^2
  +\frac{3-\hat{\beta}^4}{2\hat{\beta}}
   {\rm ln}\left(\frac{1+\hat{\beta}}{1-\hat{\beta}}\right)\right],
   \nonumber\\
   \sigma_{A_1}(\hat{s})
&=&\frac{4\pi\alpha^2Q^4_tN_c\hat{\beta}}{\hat{s}}
   \frac{(1-\hat{\beta}^2)}{4}
   \left[1+\frac{1-\hat{\beta}^2}{2\hat{\beta}}
   {\rm ln}\left(\frac{1+\hat{\beta}}{1-\hat{\beta}}\right)\right],
    \nonumber\\
   \sigma_{A_2}(\hat{s})
&=&\frac{4\pi\alpha^2Q^4_tN_c\hat{\beta}}{\hat{s}}
   \frac{1}{2}
   \left[1+\frac{1-\hat{\beta}^2}{2\hat{\beta}}
   {\rm ln}\left(\frac{1+\hat{\beta}}{1-\hat{\beta}}\right)\right].
\label{approx}
\end{eqnarray}
The expression of $\sigma_0(\hat{s})$ in Eq.~(\ref{approx})
coincides with that of Ref.~\cite{KMS}.
The $e$-$\gamma$ conversion coefficient, $\kappa$,
in Eq.~(\ref{Lum_rel}) is assumed to be $1$ in the following analysis.
$\chi$ is in the $e^+e^-$ $(e^-e^-)$ c.m. frame
the angle between the directions of maximum linear polarization
of the two laser beams, which are assumed to be
approaching nearly head-on towards the electron and positron beams,
respectively. And $L_{ee}$ is the integrated
$e^+e^-$  $(e^-e^-)$ luminosity.

The last term in Eq.~(\ref{Number}) depends
on $\sin(2\chi)$ such that the EDM contribution can be isolated
by taking the difference of events rates at $\chi=\frac{\pi}{4}$
and $\chi=-\frac{\pi}{4}$. Then one measure of the significance of
the EDM contribution is the number of standard deviations
($N_{\rm SD}$) by which
$|N(\chi=\frac{\pi}{4})-N(\chi=-\frac{\pi}{4})|$ exceeds
the expected uncertainty of the background distribution:
\begin{eqnarray}
N_{\rm SD}={\rm Re}(\tilde{d}_t)
           \frac{\sqrt{2}N_{A_2}}{\sqrt{N_0}},
\end{eqnarray}
where
\begin{eqnarray}
N_0=L_{ee}\int^{\tau_m}_{4m_t^2/s}\langle \phi_0\phi_0\rangle_\tau
    \sigma_0(\tau s),\qquad
N_{A_2}=L_{ee}\int^{\tau_m}_{4m_t^2/s}\langle \phi_3\phi_3\rangle_\tau
    \sigma_{A_2}(\tau s),
\end{eqnarray}
with $\tau_m=y^2_m$ in Eq.~(\ref{ymax}).
At the 1-$\sigma$ level the allowed maximum value of
the EDM parameter, ${\rm Re}(\tilde{d}_t)$, is
\begin{eqnarray}
|{\rm Re}(\tilde{d}_t)|_m=\sqrt{\frac{1}{2N_0}}
                                \frac{N_0}{N_{A_2}},
\label{EDM_m}
\end{eqnarray}
if no asymmetry is found. The $N_{SD}$-$\sigma$ level upper bound
is obtained simply by multiplying $|{\rm Re}(\tilde{d}_t)|_m$
by $N_{\rm SD}$.
We shall evaluate the EDM upper bound,
$|{\rm Re}(\tilde{d}_t)|_m$
for a yearly integrated luminosity of $20$ ${\rm fb}^{-1}$.

A crucial experimental issue is the optimal means for
maximizing $N_{A_2}/\sqrt{N_0}$.
It requires obtaining the smallest possible value of $x$
to make the linear polarization transfer
as large as possible. However, the large top-quark mass does
not allow $x$ to be very small. For a given c.m. energy squared, $s$,
the allowed range for $x$ is given by
\begin{eqnarray}
\frac{2m_t}{\sqrt{s}-2m_t}\leq x \leq 2(1+\sqrt{2}).
\end{eqnarray}
Numerically, for $m_t=175$ GeV the minimum value of $x$ is
$7/3$ at $\sqrt{s}=500$ GeV and $7/13$ at
$\sqrt{s}=1$ TeV, respectively. The EDM upper bound
(\ref{EDM_m}) will be very sensitive to the value of $\sqrt{s}$
since a small $x$ requires a large $s$.
Fig.~1 shows the cross section $\sigma_0$ and
the ratio $\sigma_{A_2}/\sigma_0$ with respect to
the $\gamma\gamma$ c.m. energy, $\sqrt{\hat{s}}$, for
$m_t=150$, $160$, $170$, $180$, and $190$ GeV.
The cross section is maximal near $\sqrt{\hat{s}}= 2m_t+ 100$ GeV,
while the ratio $\sigma_{A_2}/\sigma_0$ is maximal at
the threshold.
Therefore, it is important to adjust the laser energy and
electron/positron beam energies so as to obtain as many $t\bar{t}$
events as possible while retaining $x$ small so as to have
large transverse polarization of the back-scattered photons.
We make a numerical analysis for $\sqrt{s}=500$ GeV
and $m_t=150$ GeV as well as for $m_t=170$ GeV;
this facilitates direct comparison with the results of Ref.~\cite{BN}.
Table~1 shows  the $x$ dependence of the number of $t\bar{t}$
events, the ratio $N_{A_2}/N_0$, and the EDM upper bound.
$N_0$ increases while $N_{A_2}/N_0$ decreases with $x$.
As a result, the optimal upper bound $|{\rm Re}(\tilde{\mu})|_m$
for $\sqrt{s}=500$ GeV is obtained near $x=2.2$ for
$m_t=150$ GeV as
\begin{eqnarray}
|{\rm Re}(\tilde{\mu})|_m=\frac{eQ_t|{\rm Re}(\tilde{d})|_m}{2m_t}
 =5.8\times 10^{-18} ({\rm e\cdot cm}).
\label{bound}
\end{eqnarray}

The 1-$\sigma$ upper bound (\ref{bound}) should be compared with
the corresponding bound
\begin{eqnarray}
|{\rm Re}(\tilde{\mu})|_m=7.3\times 10^{-18} ({\rm e\cdot cm}),
\label{bound3}
\end{eqnarray}
which is obtained from the spin correlation study\cite{BN}
of the cascade process,
$e^+e^-\rightarrow t\bar{t}$,
$t\rightarrow bW^+(\bar{t}\rightarrow \bar{b}W^-)$, and
$W^-\rightarrow l\nu_l(W^+\rightarrow l^+\bar{\nu}_l)$\footnote{The
$l^+l^-$ channel gives the strongest bound since charged
leptons analyze the $t$ and $\bar{t}$ spins most efficiently
when the decays are as predicted by the SM.}
for $1.3\times 10^4$ $t\bar{t}$ events expected at
$\sqrt{s}=500$ GeV.
It is now clear from the bounds (\ref{bound})
and (\ref{bound3}) that, with the $\gamma\gamma$
luminosity comparable to the $e^+e^-$ luminosity,
a counting experiment of top-quark pair production events
by using linearly polarized laser beams at a photon linear collider
can measure the top-quark
EDM as accurately as or more accurately than a perfect detector
can achieve by studying the $t\bar{t}$ decay correlations
in $e^+e^-$ collisions. It is clear that by identifying the
$t$ or $\bar{t}$ mometum, we can make use of $J\geq 2$
amplitudes which are large at $\hat{s}\gg m^2_t$\cite{Choi}.

In summary,  it is possible to measure the top-quark EDM in the
polarized $\gamma\gamma$ mode by counting $t\bar{t}$ pair
production events in a straightforward manner.
In the $e^+e^-$ mode it is possible to measure the top-quark
EDM by studying $t\bar{t}$ decay correlations. Comparison
of our results for the $\gamma\gamma$ mode with previous
analyses in the $e^+e^-$ mode shows that the two approaches
are competitive.
As no information on the momenta and polarization of the top-quark
decay products is directly required, linearly-polarized
laser beams with an adjustable beam energy provide us
with a very efficient way of probing the top-quark EDM
at a photon linear collider.

\section*{Acknowledgements}

The authors would like to thank R.~Szalapski for careful
reading the manuscript.
This work was supported in part by the Japan Society for
the Promotion of Science (No. 94024) and also by the
Grant-in-Aid for Scientific
Research from the Japanese Ministry of Education, Science and
Culture (No. 05228104).

\newpage

\newcommand{\prd}[1]{Phys.~Rev.~D{#1}}
\newcommand{\prl}[1]{Phys.~Rev.~Lett.~{#1}}
\newcommand{\plb}[1]{Phys.~Lett.~B{#1}}
\newcommand{\npb}[1]{Nucl.~Phys.~B{#1}}
\newcommand{\zpc}[1]{Z.~Phys.~C{#1}}

\section*{References}

\newpage

\section*{Tables}
\renewcommand{\labelenumi}{\bf Table {\arabic {enumi}} \\}
\begin{enumerate}
\vspace*{0.5cm}
\item{The ratio $N_{A_2}/N_0$, the number of $t\bar{t}$ events
      $N_0$, and the upper EDM bound, $[{\rm Re}(\tilde{\mu}_t)]_m$,
      for selected values of $x$ with $\sqrt{s}=500$ GeV,
      $m_t=150$ GeV and $L_{ee}=20$ ${\rm fb}^{-1}$. The numbers
      in parentheses are for $m_t=170$ GeV.}

\vspace*{0.5cm}
\begin{center}
\begin{tabular}{|c|c|c|c|} \hline
 \hskip 0.2cm    $x\ \ $\hskip 0.4cm
& $\ \ N_0\ \ $
& $\ \ N_{A_2}/N_0\ \ $
& $|{\rm Re}(\tilde{\mu}_t)|_m(10^{-17} {\rm e}\cdot {\rm cm})$ \\
\hline
 $1.8$ & $156\ \ (\cdots)$
       &$\!\!\! 0.33\ \ (\cdots)$&$\!\!\! 0.76\ \ (\cdots)$\\
 $2.0$ & $395\ \ (\cdots)$i
       &$\!\!\! 0.26\ \ (\cdots)$&$\!\!\! 0.60\ \ (\cdots)$\\
 $2.2$ & $\!\!\! 680\ \ (2)$ & $0.20\ \ (0.31)$  &  $0.58\ \ (7.24)$\\
 $2.4$ & $983\ \ (36)$ & $0.16\ \ (0.26)$  &  $0.60\ \ (2.01)$\\
 $2.6$ & $1290\ \ (108)$ & $0.13\ \ (0.21)$  &  $0.65\ \ (1.39)$\\
 $2.8$ & $1594\ \ (206)$ & $0.11\ \ (0.18)$  &  $0.71\ \ (1.21)$\\
 $3.0$ & $1890\ \ (320)$ & $0.09\ \ (0.15)$  &  $0.79\ \ (1.16)$\\
 $3.2$ & $2178\ \ (445)$ & $0.08\ \ (0.13)$  &  $0.88\ \ (1.16)$\\
 $3.4$ & $2456\ \ (574)$ & $0.06\ \ (0.11)$  &  $0.98\ \ (1.20)$\\
 $3.6$ & $2724\ \ (707)$ & $0.05\ \ (0.09)$  &  $1.09\ \ (1.27)$\\
\hline
\end{tabular}
\end{center}
\end{enumerate}

\vskip 1.2cm
\section*{Figures}

\begin{enumerate}
\vspace*{0.5cm}
\item[{\bf Fig.~1}]
The cross section $\sigma_0(\hat{s})$ and the ratio
$\sigma_{A_2}(\hat{s})/\sigma_0(\hat{s})$
with respect to the $\gamma\gamma$ c.m energy
$\sqrt{\hat{s}}$ for $m_t=150$, $160$, $170$, $180$, and $190$ GeV.

\end{enumerate}

\newpage
\pagestyle{empty}
\begin{figure}[p]
\epsfxsize=6.5in
\epsfbox[51 167 470 755]{fig.ps}
\end{figure}
\end{document}